\documentclass[aps,prl,superscriptaddress,twocolumn]{revtex4}
\usepackage{color}   
\usepackage{graphicx}
\usepackage{amsmath}
\usepackage{amssymb}
\usepackage{epstopdf}
\usepackage{float}

\begin{document}
\title
{Photocatalytic activity of exfoliated graphite-TiO$_2$ nanocomposites}
\author{G. Guidetti$^{1}$, E.A.A. Pogna$^{2,3}$, L. Lombardi$^{4}$, F. Tomarchio$^{4}$, I. Polishchuk$^{5}$, R. R. M. Joosten$^{6}$, A. Ianiro$^{6}$, G. Soavi$^{4}\footnote{present address: Institut f{\"u}r Festk{\"o}rperphysik, Friedrich Schiller Universit{\"a}t Jena, 07743 Jena, Germany}$, N. A. J. M. Sommerdijk$^{6}$, H. Friedrich$^{6}$, B. Pokroy$^{5}$, A. K. Ott$^{4}$, M. Goisis$^{7}$, F. Zerbetto$^{1}$, G. Falini$^{1}$, M. Calvaresi$^{1}$, A. C. Ferrari$^{4}$, G. Cerullo$^{2}$ and M. Montalti}
\email{marco.montalti2@unibo.it}
\affiliation{Department of Chemistry G.Ciamician, Universit$\acute{a}$ di Bologna, Bologna, 40126, Italy\\
$^2$ IFN-CNR,Department of Physics, Politecnico di Milano, Milano, 20133, Italy\\
$^3$ NEST, Istituto Nanoscienze-CNR and Scuola Normale Superiore, P.zza S. Silvestro 12, I-56127, Pisa, Italy\\
$^4$ Cambridge Graphene Centre, University of Cambridge, Cambridge CB3 0FA, UK\\
$^5$ Materials Science and Engineering Department Israel Institute of Technology, echnion, Haifa, 3200003, Israel\\
$^6$ Department of Chemical Engineering and Chemistry, Eindhoven University of Technology, Eindhoven, 5612 AZ, Netherlands\\
$^7$ Global Product Innovation Department, Italcementi,  Heidelberg Cement Group, Bergamo, 24126, Italy\\}

\begin{abstract}
We investigate the photocatalytic performance of nanocomposites prepared in a one-step process by liquid-phase exfoliation of graphite in the presence of TiO$_2$ nanoparticles (NPs) at atmospheric pressure and in water, without heating or adding any surfactant, and starting from low-cost commercial reagents. The nanocomposites show enhanced photocatalytic activity, degrading up to 40$\%$ more pollutants with respect to the starting TiO$_2$-NPs. In order to understand the photo-physical mechanisms underlying this enhancement, we investigate the photo-generation of reactive species (trapped holes and electrons) by ultrafast transient absorption spectroscopy. We observe an electron transfer process from TiO$_2$ to the graphite flakes within the first picoseconds of the relaxation dynamics, which causes the decrease of the charge recombination rate, and increases the efficiency of the reactive species photo-production.
\end{abstract}
\maketitle
\section{Introduction}
Air and water pollution are major environmental risks to human health\cite{who}. According to the World Health Organization (WHO)\cite{who}, in the last decade one out of every nine deaths was related to air pollution\cite{pollution}, while at least 1.8bn people used a contaminated drinking-water source\cite{pollution2}. For air pollution remediation, environmental contaminants\cite{Hoffmann1995} (e.g. NO, NO$_2$, SO$_2$, suspended organic particulate, volatile organic compounds, aromatic hydrocarbons, etc.) must be turned into harmless compounds. This can be achieved exploiting photocatalysts to absorb light and produce reactive holes (h) and electrons (e) that degrade the pollutants via redox processes\cite{Hirakawa_1}. The photocatalytic quantum efficiency (PQE, adimensional) is defined as the ratio between the rate at which the target molecules undergo photo-degradation (moles of molecules per unit time) [mol s$^{-1}$], and the rate of photon absorption (moles of absorbed photons per unit time) [mol s$^{-1}$]\cite{Serpone_1,Serpone_2}. Since photocatalytic degradation relies on the Sun and on the photocatalyst, not consumed during the process\cite{Hoffmann1995,ZhangQ2015}, this is a potentially low-cost and environment friendly approach for pollution abatement\cite{Hoffmann1995}.

Amongst oxide semiconductor photocatalysts\cite{Hoffmann1995} (such as ZnO, FeO$_3$, WO$_3$), titanium dioxide nanoparticles (TiO$_2$-NPs) have a wide range of applications, including self-cleaning\cite{Lai2016}, sterilization of surfaces\cite{Nishimoto2013}, air\cite{Mamaghani2017} and water\cite{Yu2016} purification. TiO$_2$-NPs have the advantages of stability in water\cite{Hoffmann1995}, non-toxicity\cite{Xia2006} and low cost ($\sim$1900USD/Ton at 2016 prices\cite{TiO2cost}). Due to its wide band gap (3.25 eV\cite{Hoffmann1995}), TiO$_2$ absorbs only the UV part of the solar spectrum\cite{Linsebigler1995}. TiO$_2$-NPs with diameter$>$10nm\cite{Serpone_2} do not display quantum confinement effects, which would result in a blue shift of the absorption spectra\cite{Serpone_1}. Hence, TiO$_2$-NPs exploit just the UV part ($\sim$4$\%$\cite{Foyomoreno2003}) of the solar radiation to perform photodegradation\cite{Hoffmann1995,ChenX2007,Fujishima2008,Linsebigler1995}, wasting$\sim$96$\%$ of the usable spectrum. Even considering only the UV component, TiO$_2$-NPs have a modest PQE$\sim$10$\%$\cite{Serpone_2}, limited by the recombination of the photo-generated e-h pairs that occurs with 90$\%$ quantum efficiency\cite{Serpone1997}. The PQE increases with the number of generated e-h pairs per absorbed photon, i.e. the photo-generation yield\cite{Serpone1997}, and with the carriers' lifetime\cite{Serpone_2}. Integration with materials able to accept e or h may slow down charge recombination, leading to a PQE increase.

The integration of TiO$_2$ with carbon materials, such as nanotubes\cite{Dai2009}, dots\cite{Sun2014}, graphene oxide (GO)\cite{Reddy2015} and reduced graphene oxide (RGO)\cite{Han2012}, was pursued to enhance PQE\cite{Reddy2015,Han2012,Leary2011,Tu2013,WangH2013,ZhangN2015,ZhangN2012,Chen2010,Du2011,Gao2012,Guo2011,Jiang2011,KimC2012,KimI2011,Li2011,Liang2010,Liang2011,Liang2012,Liu2011,LiuS2013,Pan2012,
Pastrana-Martinez2012,Shah2012,Shen2011,Thuy-Duong2011,WangP2013,WangY2010,Williams2008,YangN2013,Zhang2010,ZhangJ2011,ZhangL2008,ZhangP2011,ZhangY2010,ZhangY2011,ZhangY2012,Zhao2012,Yang2010,Yeh2013,
Morais2016}. The e-h pair generation and evolution in TiO$_2$/carbon composites, such as TiO$_2$/RGO\cite{Wang_1,Manga_1,Morais2016} and TiO$_2$/graphene quantum dots (GQD)\cite{Kenrick_1}, was investigated by transient absorption (TA) spectroscopy\cite{DeSilvestri2017}. In TiO$_2$/GQD\cite{Kenrick_1}, TiO$_2$/RGO\cite{Wang_1} and Ti$_{0.91}$O$_2$/RGO\cite{Manga_1}, TiO$_2$ acts as e acceptor when excited with visible light below the TiO$_2$ optical gap. In TiO$_2$/GQD, the e-injection occurs with a time constant$<$15fs\cite{Wang_1}. In Ti$_{0.91}$O$_2$/RGO with 0.1 wt$\%$ RGO, RGO was found to act as e acceptor, decreasing the recombination rate in TiO$_2$\cite{Morais2016}. Thus, when excited with UV photons above the TiO$_2$ gap, RGO acts as e-acceptor causing the decrease of the charge carriers' recombination rate, resulting in PQE enhancement. However, Ref.\onlinecite{Morais2016} did not quantify the lifetime of the photo-generated carriers, because of the limited time resolution used ($\sim \mu$s\cite{Schneider2018}). Ref.\onlinecite{Long_1} theoretically investigated the charge transfer processes, predicting that charge and energy transfer in TiO$_2$/single layer graphene (SLG) would proceed in both directions, depending on the energy of the excited charges. Here, we apply ultrafast transient absorption spectroscopy to investigate charge separation in exfoliated graphite/TiO$_2$. Our results explain the mechanism responsible for the increased PQE in TiO$_2$/carbon composites.

Our TiO$_2$/exfoliated graphite (TiO$_2$/Gr) photocatalyst is prepared by sonication-assisted exfoliation of graphite in presence of TiO$_2$-NPs, using commercial starting materials suitable for large scale production. Liquid-phase exfoliation (LPE) of graphite typically exploits surfactants\cite{Hernandez2008,BonaMatToday}, such as sodium deoxycholate\cite{Lotya2010} and pluronics\cite{Nanoscale2015}. Here we use the TiO$_2$-NPs themselves to exfoliate graphite in water and produce the photocatalytic composite. The exfoliation process is investigated varying both sonication time and concentration of TiO$_2$-NPs and comparing the chemical composition and crystal structure by high-resolution powder X-Ray diffraction (HR-PXRD). The photocatalytic activity is evaluated by measuring the rate of degradation of a model organic compound (Rhodamine B) in water under UV irradiation. An increase up to$\sim$100$\%$ of the degradation rate, with respect to the TiO$_2$-NPs used as starting material, is observed. The photophysical mechanism underlying this enhanced photocatalytic activity is investigated by ultrafast TA spectroscopy with sub-200fs time resolution and broad spectral coverage (430-1400nm). We compare the decays of photo-generated e-h pairs in the composite with those in pristine TiO$_2$ and we observe that TiO$_2$-NPs inject e into the graphite flakes. The increased photo-production of reactive species explains the photocatalytic activity improvement, with exfoliated graphite acting as e-acceptor.
\section{\label{Disc}Results and Discussion}
The composites are prepared via ultra sonication of graphite in a 2mg/ml aqueous dispersion of TiO$_2$-NPs for 4 hours (ELMATransonicT460/H-35kHz) at 40 $^{\circ}$C. The exfoliation is performed in Millipore ultrapure water (resistivity 18.2 M$\Omega$$\cdot$cm at 25$^{\circ}$C). We use flakes from Sigma-Aldrich with size$\sim$150$\mu$m and TiO$_2$-NPs in the anatase form from HOMBIKAT AHP 200, Sachtleben Chemie GmbH (purity of the crystalline phase$\geq$94$\%$w/w, average surface area$\sim$193m$^{2}$/g). To explore a wide range of conditions, two TiO$_2$-graphite mass ratios are used: 1:1 in TiO$_2$-Gr1:1 and 10:1 in TiO$_2$-Gr10:1.

In order to study the effect of TiO$_2$-NPs during liquid-phase sonication, we perform HR-PXRD measurements as a function of sonication time. Samples are loaded into 1mm borosilicate glass capillaries and diffraction patterns collected at ambient temperature with an incident X-ray wavelength of 0.319902\AA. The full width at half maximum (FWHM) of the $\{002\}$ graphite diffraction peak is deduced by the Rietveld refinement method\cite{Young_1}, using the General Structure Analysis System (GSAS) program and EXPGUI interface\cite{Tanaike1997,Fitch2004}.

\begin{figure}
\centerline{\includegraphics[width=90mm]{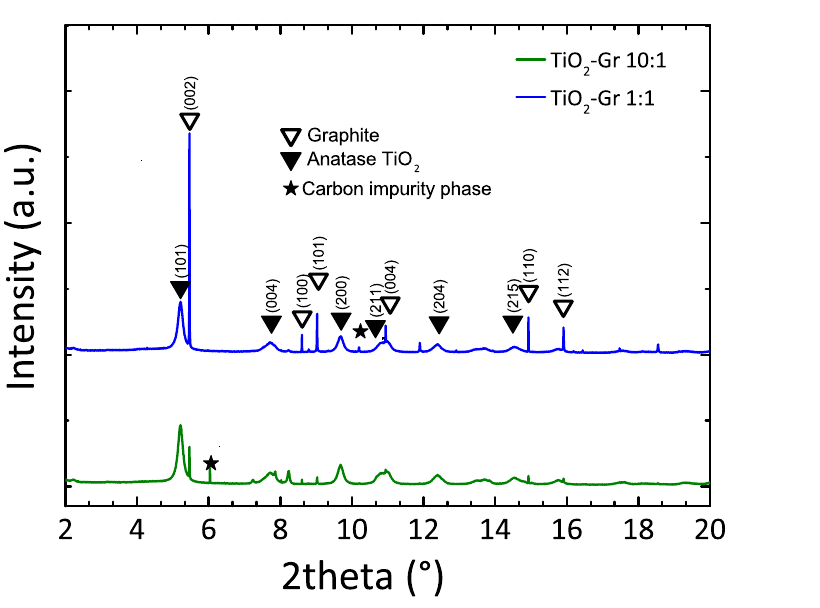}}
\caption{HR-PXRD diffraction profiles of TiO$_2$/Gr1:1 and 10:1 sonicated for 30 mins, with peak assignment}
\label{fig:Fig3}
\end{figure}

Diffraction patterns collected after 30min sonication are shown in Fig.\ref{fig:Fig3}. For longer time, up to 4h, further structural changes are not observed. Thus samples sonicated for 30min can be considered as the final products. Fig.\ref{fig:Fig3} confirms the presence of TiO$_2$ anatase (as for The Joint Committee on Powder Diffraction Standards, JCPDS 21-1272\cite{bookJC}) and graphite (JCPDS 75-2078)\cite{Toby2001}. Moreover, the basal reflection shifts towards higher d-spacings (d002=3.357\AA) with respect to graphite (JCPDS 75-2078, d002=3.347\AA\cite{Toby2001}). This suggests that TiO$_2$-NPs assisted exfoliation increases the interplanar spacing of the resulting flakes. The $\{002\}$ diffraction peak of TiO$_2$-Gr10:1 has lower intensity than in TiO$_2$-Gr1:1. This indicates that an increase in TiO$_2$-NPs concentration leads to a decrease in the number of planes oriented along $\{002\}$\cite{Toby2001}. As the concentration of TiO$_2$-NPs increases, the {002} reflection broadens and the corresponding FWHM increases. The broadened FWHM is due to a smaller crystallite size\cite{Toby2001,Holcomb1973} ($\sim$195nm, $\sim$241nm and $\sim$255nm for TiO$_2$-Gr10:1, TiO$_2$-Gr1:1 and graphite, respectively) as determined by the Rietveld method\cite{Young_1} using the software GSAS-II of Ref.\onlinecite{Toby2013}.
\begin{figure}
\centerline{\includegraphics[width=90mm]{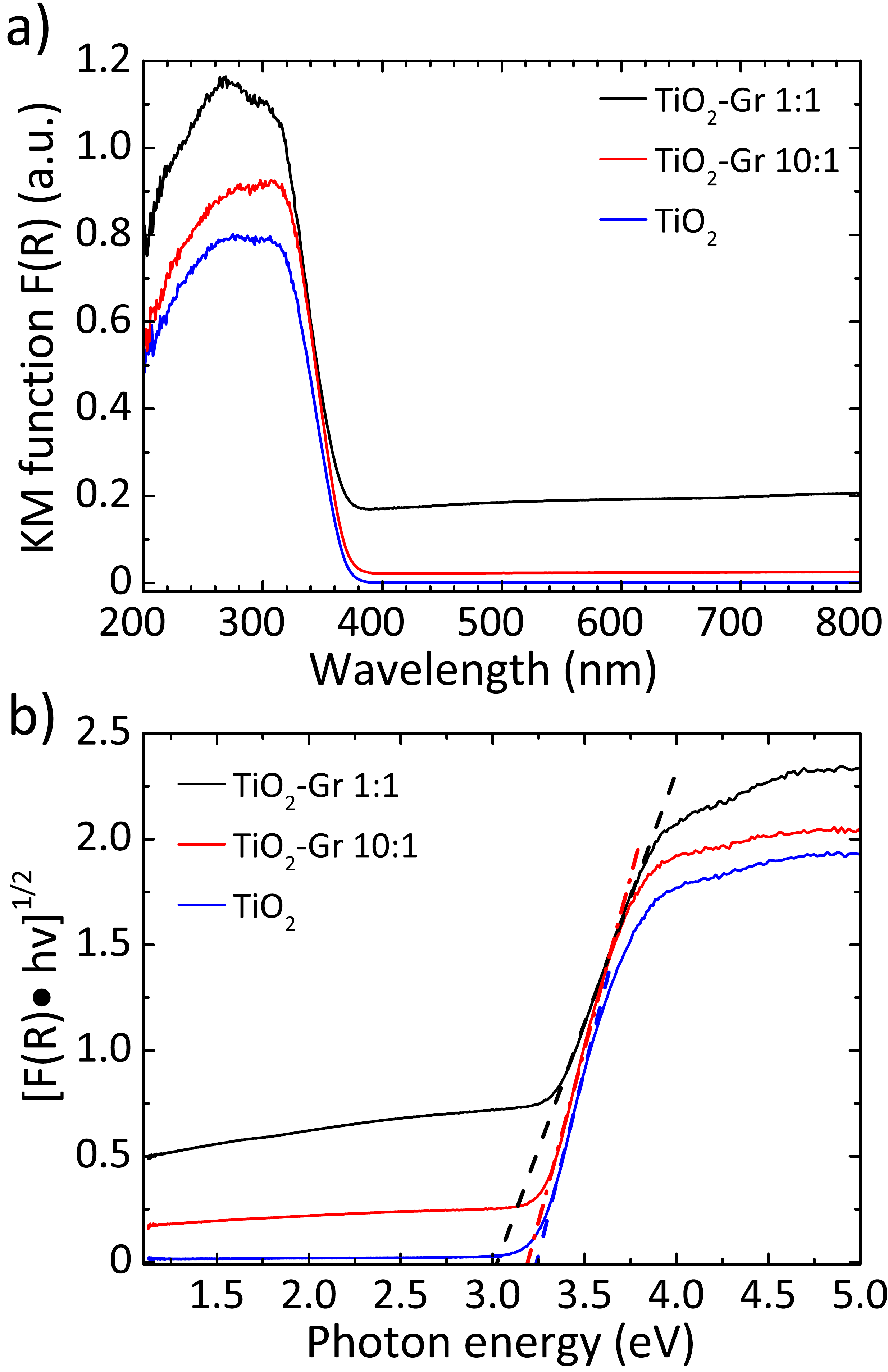}}
\caption{a) \textit{F(R)} elaborated and normalized with KM\cite{Kubelka1931}; b) Tauc plot of the modified \textit{KM} function. Dotted lines show the linear extrapolation of the Tauc gap.}
\label{fig:Fig6}
\end{figure}

We then investigate the photo-physical properties of the samples by UV-visible (UV-Vis) diffuse reflectance spectrometry with a Perkin Elmer Lambda45 UV-Vis spectrophotometer with Harricks praying mantis diffuse reflectance. For each sample, 10mg is mixed with a 500mg NaCl matrix. We use a quartz cuvette with 0.5cm optical path. The reflectance background of NaCl (reference) is taken as baseline for each measurement. The diffuse reflectance can be linked to the absorption coefficient through the Kubelka-Munk\textit(KM) function\cite{Kubelka1931}\textit{F(R)}. For a sample thickness$>$3mm\cite{Kortum2012,Workman1998}, with no light transmission, F(R) can be written as\cite{Christy1995}:
\begin{equation}
F(R)= (1-R)^2/2R=K/s=2.303 \epsilon\cdot c/s
\label{eqa1}
\end{equation}
where \textit{R} is the absolute reflectance, \textit{K} [cm$^{-1}$] is the absorption coefficient, \textit{s} [cm$^{-1}$] is the scattering coefficient, $\epsilon$ is the absorptivity [$mol\cdot$L$^{-1}\cdot$cm$^{-1}$] and \textit{c} is the concentration [M]. Since the samples are dispersed into a non-absorbing matrix (NaCl), \textit{s} in Eq(\ref{eqa1}) can be assumed to be that of NaCl and constant\cite{Jackson_1}. As a consequence, \textit{F(R)} is proportional to \textit{K}. Fig.\ref{fig:Fig6}a plots the spectra of pristine TiO$_2$, TiO$_2$-Gr10:1 and TiO$_2$-Gr1:1. A transition from the valence to the conduction band of TiO$_2$ can be seen at$\sim$340-360nm in all samples, as expected for anatase based composites\cite{Reddy_1,Murphy_1}. The presence of exfoliated graphite gives rise to absorption from 400 to 800nm\cite{Smausz2017}, and \textit{F(R)} is higher with respect to pristine TiO$_2$. An estimation of the band gap can be obtained applying the Tauc equation, which relates absorption edge, energy of incident photons $h\nu$ and Tauc gap $E_T$\cite{Ngamta2013}:
\begin{equation}
K h\nu=A(h\nu-E_T)^n
\label{eqTauc}
\end{equation}
where A is a proportionality constant and the index n depends on the interband transitions dominating the absorption. In TiO$_2$ n=2 is applied\cite{Ngamta2013} because the interband transitions are indirect. E$_T$ can be determined by a linear extrapolation of (F(R)h$\nu$)$^{1/2}$ versus h$\nu$, Fig.\ref{fig:Fig6}. We get E$_T\sim3.25eV$ for pristine TiO$_2$ decreasing to $\sim3.20$ and $\sim3.02eV$ for TiO$_2$-Gr10:1 and TiO$_2$-Gr1:1.
\begin{figure}
\centerline{\includegraphics[width=90mm]{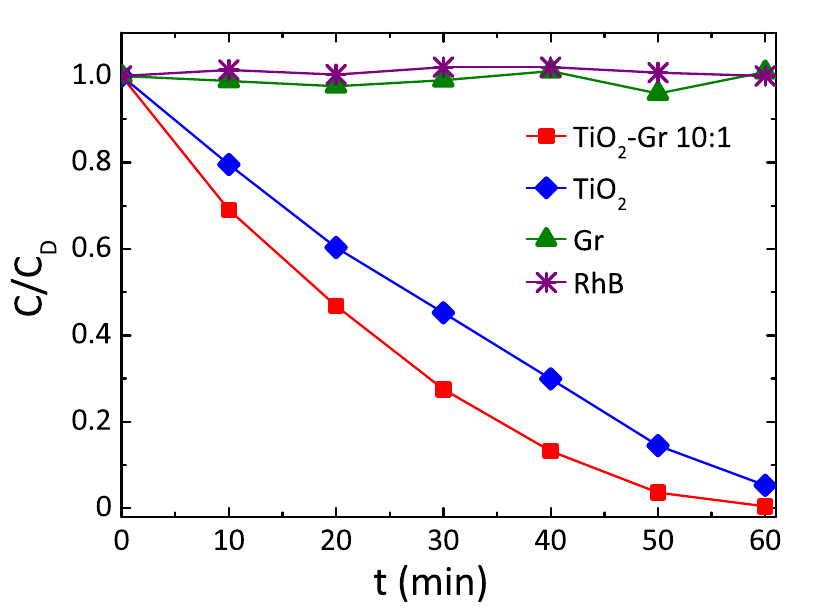}}
\caption{Photocatalytic degradation of RhB under UV light irradiation in the presence of TiO$_2$-Gr10:1, pristine TiO$_{2}$ and Gr reference.}
\label{fig:Fig8}
\end{figure}
\begin{figure}
\centerline{\includegraphics[width=90mm]{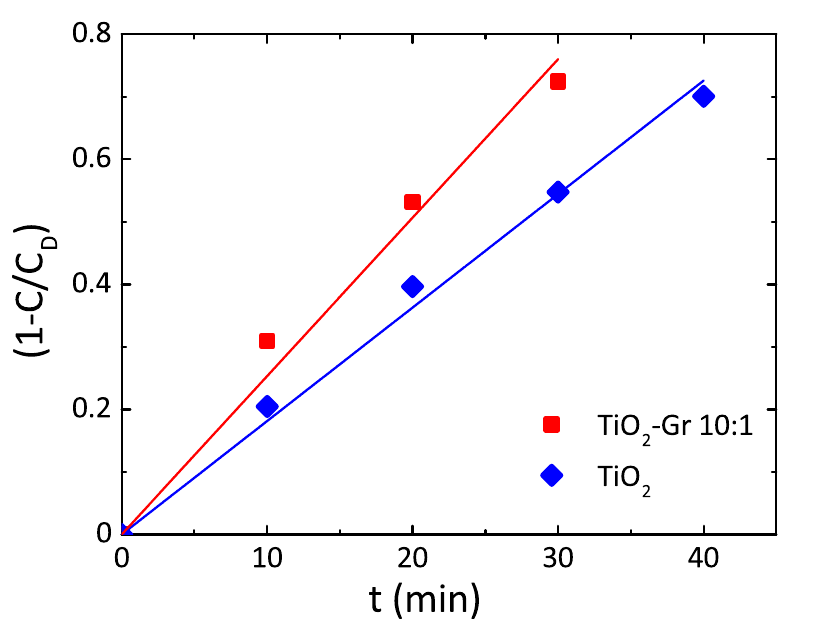}}
\caption{Photodegradation kinetics of RhB for TiO$_{2}$, TiO$_{2}$-Gr10:1. Lines are fits to the data with Eq.\ref{eq4}.}
\label{fig:Fig9}
\end{figure}
\begin{figure*}
\centerline{\includegraphics[width=180mm]{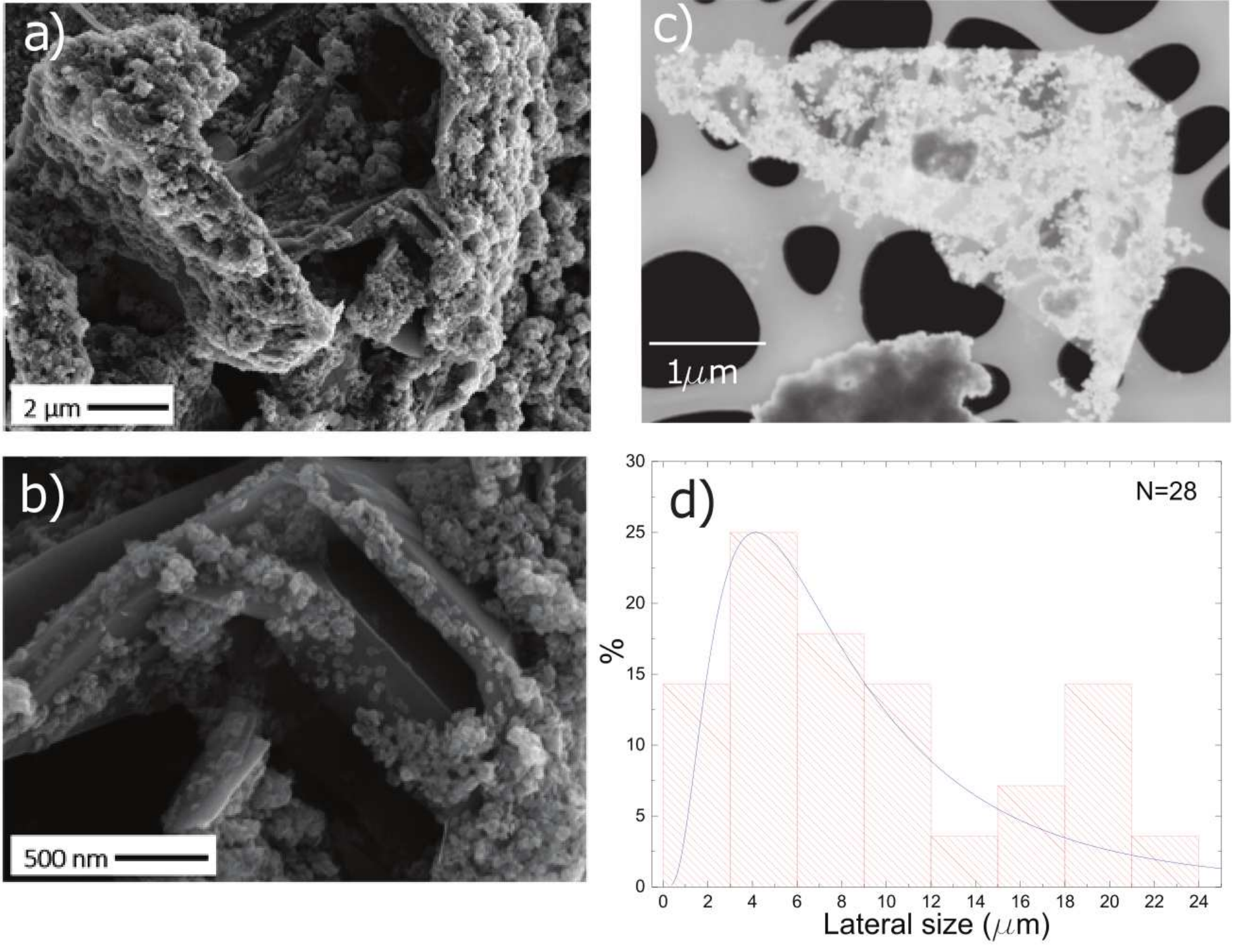}}
\caption{a) SEM image of TiO$_{2}$-Gr10:1. b) Higher magnification of a) showing flakes decorated with NPs. c) Representative STEM image of a flake in TiO$_{2}$-Gr10:1. d) Distribution of flakes lateral size as determined by STEM of N=28 flakes.}
\label{fig:Fig1}
\end{figure*}
\begin{figure}
\centerline{\includegraphics[width=90mm]{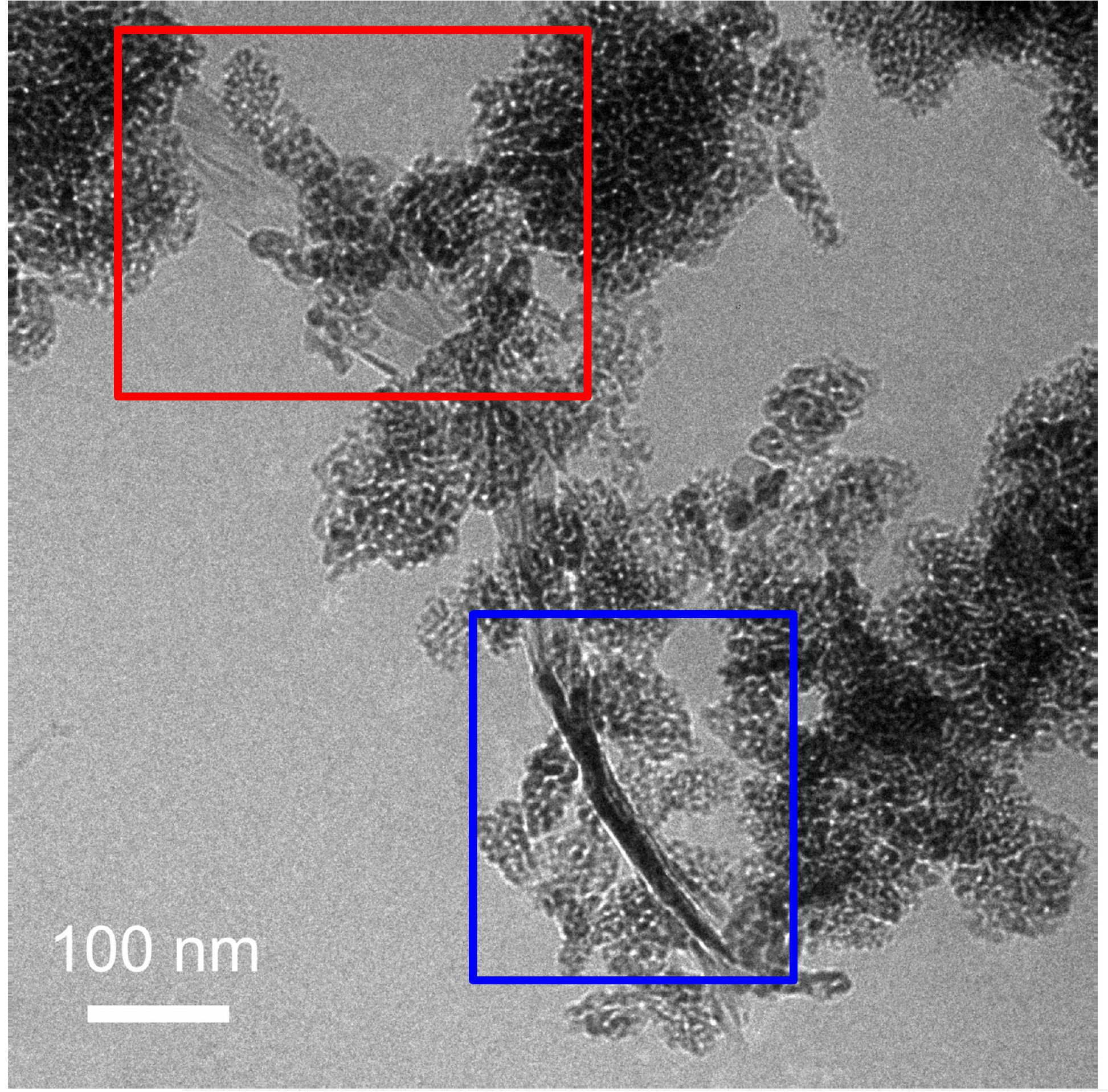}}
\caption{Cryo-TEM of TiO${_2}$-Gr10:1 in suspension with TiO${_2}$-NPs decorating the flake surface (red rectangle) and the edges (blue rectangle).}
\label{fig:Fig2}
\end{figure}
\begin{figure*}
\centerline{\includegraphics[width=180mm]{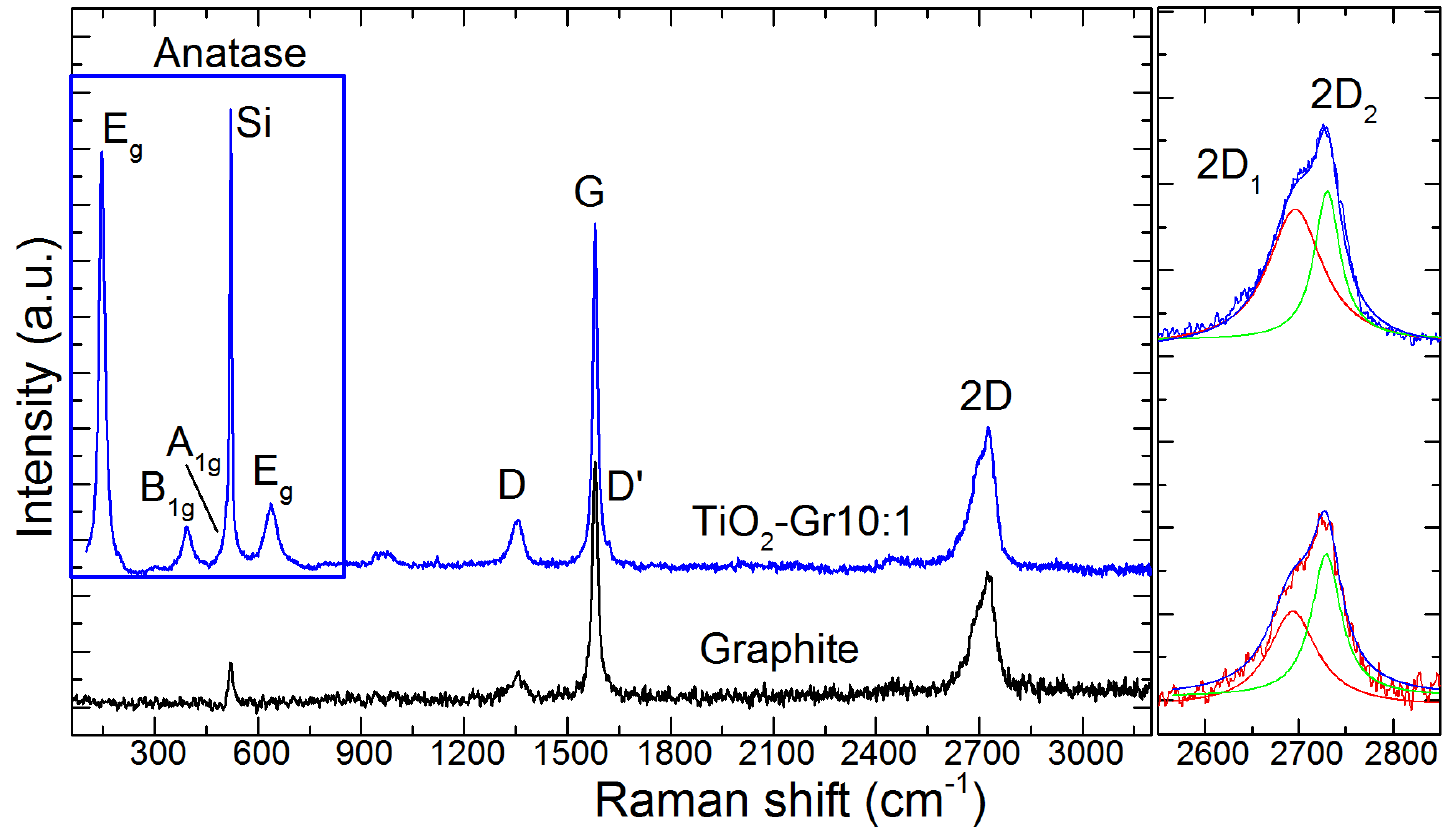}}
\caption{Representative Raman spectra at 514.5 nm for graphite (black curve) and TiO$_{2}$-Gr10:1 (blue curve) both recorded on a Si/SiO$_2$ substrate.}
\label{fig:Fig5}
\end{figure*}

The photocatalytic activity is investigated by measuring the photo-degradation of a molecular non-azo-dye (rhodamineB; RhB). This compound is taken as model for organic volatile pollutants since its molecular structure is close to that of the environmental contaminants used in industry and agriculture\cite{Jain_1}. This test follows the same procedures used to characterize other TiO$_2$-carbon composites\cite{ZhangN2012,Liang2010}. TiO$_2$, TiO$_2$-Gr10:1 and TiO$_2$-Gr1:1 are dispersed in an aqueous solution and sonicated for 4h. In order to understand the effect of the graphite flakes on PQE, the amount of TiO$_2$-Gr10:1 and TiO$_2$-Gr1:1 is chosen to guarantee the same concentration of TiO$_2$ (2mg/ml) inside each suspension. We test 10ml mixtures comprising 2.86$\%$ml of an aqueous solution of RhB (0.05mg/ml,1$\cdot$10$^{-4}$M), 2.14$\%$ ml H$_2$O and 50$\%$ suspension of TiO$_2$-Gr10:1 or TiO$_2$-Gr1:1. Considering the affinity of graphitic flakes, due to the $\pi$-$\pi$ stacking of their aromatic systems, for polycyclic aromatic and cationic compounds like RhB\cite{Guidetti2016}, the suspensions are magnetically stirred for 40min in the dark, in order to attain adsorption-desorption equilibrium between composite and dye. Ref.\onlinecite{Guidetti2016} reported that, when RhB is adsorbed onto 2-3 layers graphene flakes, there is a ground state interaction that leads to a decrement in the intensity of UV/Vis absorption and photoluminescence (PL) of the dye independent of photodegradation. It is thus necessary to determine the fraction of RhB that remains free inside the solution, since this is required to discriminate whether the change in the dye concentration under irradiation is due to a photoreaction or to adsorption. To obtain the adsorption, after stirring in the dark, 0.75ml of the RhB-composite suspension is taken and centrifuged at 9000rpm for 10min at T=25$^o$C in order to separate the sample from the RhB solution. The upper 0.5ml is collected and diluted with water (1:6 ratio) to reach the 3ml volume of analysis of a standard cuvette for a UV/Vis spectrophotometer. The concentration C$_D$ (mol L$^{-1}$) of free RhB after incubation in the dark is derived from UV/Vis absorption spectra ($\lambda$max=554nm) recorded at 25$^{\circ}$ with a Cary300 UV-Vis spectrophotometer and a 180$\mu$m path-length cuvette. The percentage \textit{Ads} of RhB adsorbed is calculated starting from the initial concentration C$_0$ (mol$\cdot$L$^{-1}$) of the used dye, as\cite{PATIL}:
\begin{equation}
Ads= [(C_0-C_D)/C_0]\cdot 100
\label{eq2}
\end{equation}
\begin{table}
\caption{Percentage of photodegraded RhB after 20 and 40mins irradiation and adsorption of RhB after incubation in the dark}
\centering
\begin{tabular}{c c c c}
\hline
& TiO$_{2}$ & TiO$_{2}$-Gr10:1 & TiO$_{2}$-Gr1:1 \\ [0.5ex]
\hline
\textit{P(20min)} & 38$\%$ & 54$\%$ & 45$\%$ \\ 
\hline
\textit{P(40min)} & 66$\%$ & 87$\%$ & 64$\%$ \\
\hline
\textit{Ads} & 5$\%$ & 5$\%$ & 35$\%$ \\
\hline
\end{tabular}
\label{table:Table1}
\end{table}
The photoreactivity after photoexcitation of TiO$_2$ is investigated by exposing each sample to a lamp emitting in the UVA/UVB range (280-400nm), matching the absorption spectra of the composites,Fig.\ref{fig:Fig6}. The lamp has irradiance, i.e. emitted power per unit area, \textit{I}$\sim$3W/m$^2$ in the UVA (280-315nm) and$\sim$13.6 W/m$^2$ in UVB (315nm-400nm), at 0.5m from the source. The samples are placed 35cm from the lamp. We use 1mW UVA/UVB for 60mins, sampling 0.75ml every time interval \textit{t} of 10mins. The collected volumes are centrifuged, diluted and analyzed with the same procedure used for the determination of C$_D$, detecting the concentration C(\textit{t}) of RhB not degraded after \textit{t} from the beginning of the irradiation. The percentage of RhB photodegraded, \textit{P (t)} is\cite{Natarajan2011}:
\begin{equation}
P(t)= [(C_D-C(t))/C_0] \cdot 100
\label{eq3}
\end{equation}
Using this approach, the photocatalytic activity is assessed independently of the possible adsorption of the dye onto the surface of the photoactive material, since the concentration of the dye after pre-equilibration is taken as a reference. For TiO$_2$-Gr10:1, Table \ref{table:Table1} shows an increment of \textit{P(t)} with respect to TiO$_2$ of$\sim$16$\%$ after 20mins and$\sim$21$\%$ after 40mins. For TiO$_2$-Gr1:1, the increment is$\sim$7$\%$ after 20mins while a decrement$\sim$2$\%$ occurs after 40mins. The adsorption of RhB increases from$\sim$5$\%$ in TiO$_2$-Gr10:1 to$\sim$35$\%$ in TiO$_2$-Gr1:1. These results indicate that TiO${_2}$-Gr1:1 does not show improvement in photocatalytic activity with respect to TiO$_{2}$. The reason for this is the presence of a residual of graphite that is not electronically interacting with TiO$_{2}$ in TiO$_{2}$-Gr1:1. This excess of graphite is demonstrated by the broad absorption in the 400-800 nm region in Fig.\ref{fig:Fig6}. This graphite  adsorbs RhB as demonstrated by the increase of the adsorbed fraction from 5$\%$ to 35$\%$ but it is not photocatalyically active. As a result, the fraction of light absorbed by this non-photochemically active component is dissipated without giving photodegradation of RhB, causing a decrease of P. We thus identify TiO$_2$-Gr10:1 as a promising photocatalytic compound since it gives an enhanced \textit{P(t)} with respect to TiO$_{2}$, for a similar RhB adsorption. The observed lack of improvement in photocathalytic activity of TiO$_2$-Gr1:1 with respect to TiO$_2$ is in agreement with Refs.\onlinecite{ZhangN2012,WangY2010,Zhang2010,Zhu2012}, where the adsorption and photocatalytic activity of TiO$_2$ composites with GO and RGO was reported: a GO/TiO$_2$ or RGO/TiO$_2$ weight$>$10$\%$ w/w was associated with a decrease of photocatalytic activity. Hence, we focus on TiO$_2$-Gr10:1 hereafter.

Fig.\ref{fig:Fig8} compares the concentration of RhB during photodegradation upon UV irradiation for: i) TiO$_2$-Gr10:1, ii) reference TiO$_2$, iii) graphite, iv) no photocatalyst. The trends indicate that the dye's degradation temporal profile is a combination of a zero-order and a first-order kinetics. In zero-order kinetics, the rate is independent of the reactant concentration and the RhB concentration decreases linearly with time\cite{Zhou2018}, while in first order, the rate is proportional to the dye concentration.

Since neither zero-order nor first order models fit the data of Fig.\ref{fig:Fig8}, we use a pseudo-zero-order kinetic model commonly adopted in the case of organic dye photodegradation in heterogeneous systems\cite{Rajeshwar_1}:
\begin{equation}
1-C(t)/C_D= kt
\label{eq4}
\end{equation}
where \textit{k}(min$^{-1}$) is the kinetic constant. Fig.\ref{fig:Fig9} fits the data with Eq.\ref{eq4}. This gives k(min$^{-1}$)$\sim$0.018 and$\sim$0.025 for TiO$_{2}$ and TiO$_{2}$-Gr10:1, again indicating that TiO$_2$-Gr10:1 has higher photoactivity than TiO$_2$.

The morphology of TiO${_2}$-Gr10:1 is investigated by scanning electron microscopy (SEM, Quanta3D, FEI Company). Fig.\ref{fig:Fig1}a shows graphitic flakes covered by TiO$_{2}$-NPs. The higher magnification image Fig.\ref{fig:Fig1}b indicates that the flakes edges are decorated by NP agglomerates. The lateral size of the flakes is evaluated by Scanning Transmission Electron Microscopy (STEM, Magellan 400L FEI) depositing$\sim$20$\mu$l TiO${_2}$-Gr10:1 on a holey carbon Cu grid (300 mesh). From a statistical analysis of isolated flakes similar to that in Fig.\ref{fig:Fig1}c, an average lateral size$\sim$5$\mu$m is estimated, Fig.\ref{fig:Fig1}d.

To exclude that the TiO${_2}$-NPs adhesion to the flakes is due to the drying of the TiO${_2}$-Gr10:1 suspension, we perform Cryo-TEM (CRyoTitan FEI) experiments. 20$\mu$l TiO$_{2}$-Gr10:1 is deposited on a holey carbon grid (Quantifoil R2/2 200mesh), then the sample is loaded into the chamber of a FEI Vitrobot$^{TM}$ Mark III, that maintains 100$\%$ humidity at 4${^o}$C. Inside the chamber there are two blotting papers on either side of the sample, which close on the grid and leave a layer of suspension$\sim$hundreds nm thick\cite{fei}. The sample is then plunged into liquid ethane at -183.3 ${^o}$C, which avoids the formation of ice crystals\cite{bib3}, creating a vitreous ice (amorphous solid form of water)\cite{bib3}. This allows us to investigate the morphology of TiO${_2}$-Gr10:1 in the liquid phase, confirming that TiO${_2}$-NPs adhere to the flakes, both on the surface (red rectangle) and at the edges (blue rectangle), Fig.\ref{fig:Fig2}.

\begin{figure}
\centerline{\includegraphics[width=90mm]{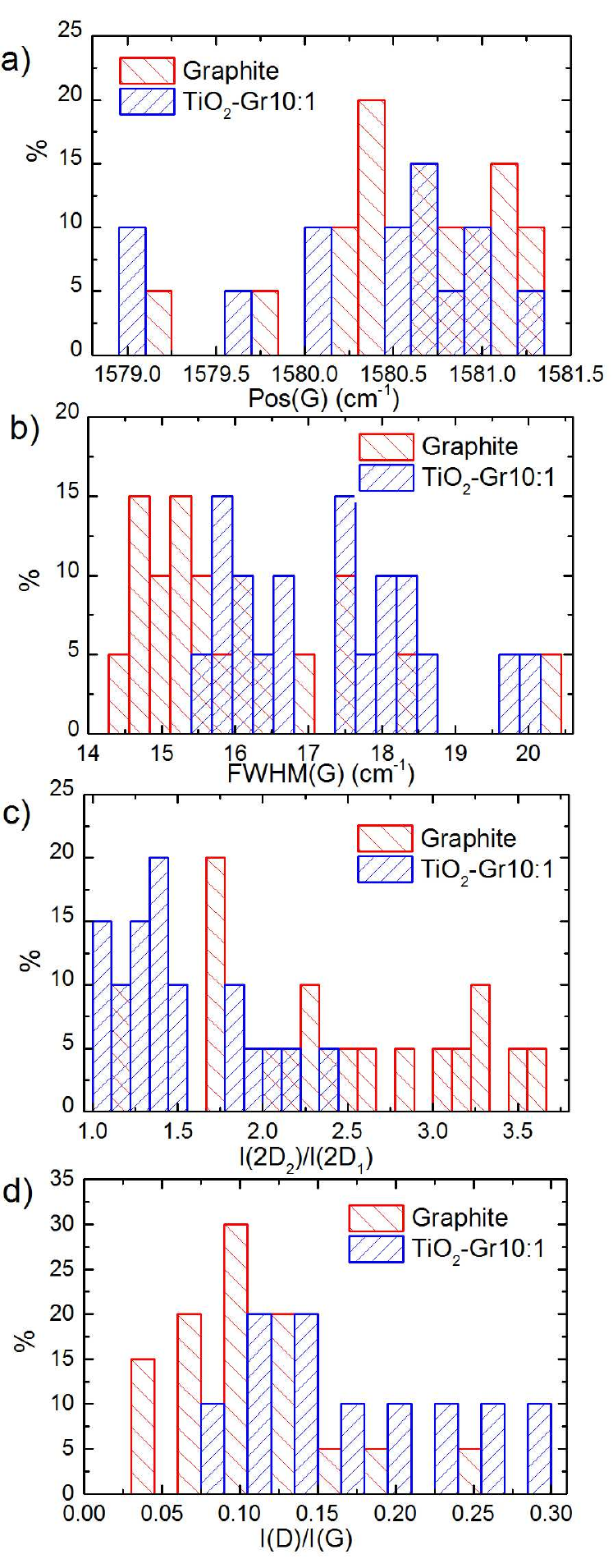}}
\caption{Distribution of: (a) Pos(G), (b) FWHM(G), (c) I(2D$_{2}$)/I(2D$_{1}$) and (d) I(D)/I(G).}
\label{fig:Raman}
\end{figure}

TiO$_2$-Gr10:1 and the starting graphite are also characterized by Raman spectroscopy. 60$\mu$l is drop cast onto a Si/SiO$_2$ substrate, then heated at 100 $^{\circ}$C for 20mins, to ensure water evaporation. Raman spectra are acquired at 514.5nm using a Renishaw InVia spectrometer with a Leica DM LM microscope and a 50x objective. The power on the sample is kept below 1mW to avoid any possible damage and heating. The spectral resolution is$\sim$1cm$^{-1}$. A statistical analysis is performed as follows: the substrate is divided into 4 regions$\sim$500$\times$500 $\mu$m$^2$ and in each 5 points are acquired. Fig.\ref{fig:Fig5} plots representative Raman spectra of the starting graphite (black line) and of TiO$_2$-Gr10:1 (blue line) both on Si/SiO$_2$. The peaks at 144, 397, 518 and 639 cm$^{-1}$ are the \textit{E$_g$}, \textit{B$_{1g}$}, \textit{A$_{1g}$} and \textit{E$_g$} modes of anatase TiO$_2$\cite{Ohsaka1978}. The TiO$_2$ peak at 518cm$^{-1}$ is very close to the first order peak of silicon $\sim$521cm$^{-1}$\cite{Temple1973} and they are partially overlapping. The crystallite size of TiO$_2$-NPs can be estimated from the position Pos(\textit{E$_g$}@144cm$^{-1}$) and FWHM(\textit{E$_g$}@144cm$^{-1}$)\cite{Swamy2005}. In our case Pos(\textit{E$_g$}@144cm$^{-1}$)$\sim$147cm$^{-1}$ and FWHM$\sim$20cm$^{-1}$ correspond to a NPs size$\sim$7nm\cite{Swamy2005}, in agreement with an estimate from TEM images, as in Fig.\ref{fig:Fig2}, of$\sim$5-10nm. Figs.\ref{fig:Raman}a,b show no significant difference between Pos(G) and FWHM(G) of graphite and TiO$_2$-Gr10:1. The 2D peak shape for TiO$_2$-Gr10:1 still resembles that of graphite\cite{FerrariPRL2006} with two components (2D$_1$, 2D$_2$), but their intensity ratio I(2D$_2$)/I(2D$_1$) is reduced from 2.4 to $\sim$1.4, Fig.\ref{fig:Raman}c. This indicates that the bulk flakes have undergone exfoliation\cite{Ferrari2000}.

When compared to the initial graphite, TiO$_2$-Gr10:1 has a higher I(D)/I(G) and FWHM(G). I(D)/I(G) varies inversely with the crystal size, L$_a$, according to the Tuinstra and Koenig (TK) equation: I(D)/I(G)$\sim$4.4nm/L$_a$\cite{Ferrari2000, Tuinstra1970}. Alternatively, this can be seen as proportional to the average interdefect distance, L$_D$: I(D)/I(G)$\sim$130nm/L$_D^2$\cite{Cancado2011}. I(D)/I(G) can also be affected by doping\cite{Bruna2014}. The lack of up shift of Pos(G) and of FWHM(G) narrowing in TiO$_2$-Gr10:1 when compared to graphite suggests a level of doping similar to the starting graphite, with a negligible effect on I(D)/I(G). We get L$_D\sim$31nm and L$_a\sim$33nm for TiO$_2$-Gr10:1 while for graphite these are$\sim$43nm and$\sim$63nm. Given the average flakes lateral size in Fig.\ref{fig:Fig2}, these numbers reflect the defective nature of the starting graphite, and show that defects increase after sonication. L$_a$ determined by Raman is consistent with that derived from HR-PXRD, although lower.
\begin{figure}
\centerline{\includegraphics[width=90mm]{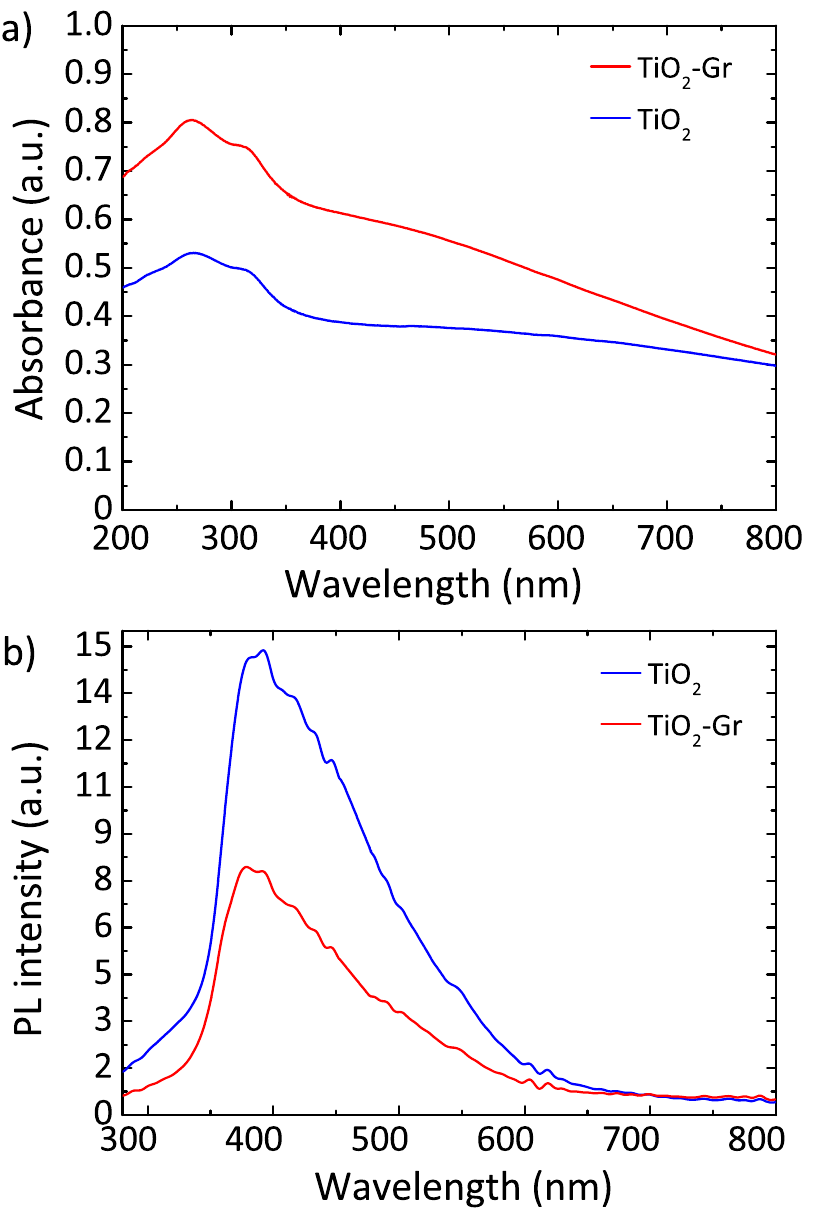}}
\caption{a) UV-Vis absorbance spectra of pristine TiO$_2$ and TiO$_2$-Gr10:1 in the 200-800nm range. b) PL spectra for 266nm excitation of TiO$_2$ and TiO$_2$-Gr10:1 in the 280-800nm range.}
\label{fig:Fig7}
\end{figure}
\begin{figure*}
\centerline{\includegraphics[width=170mm]{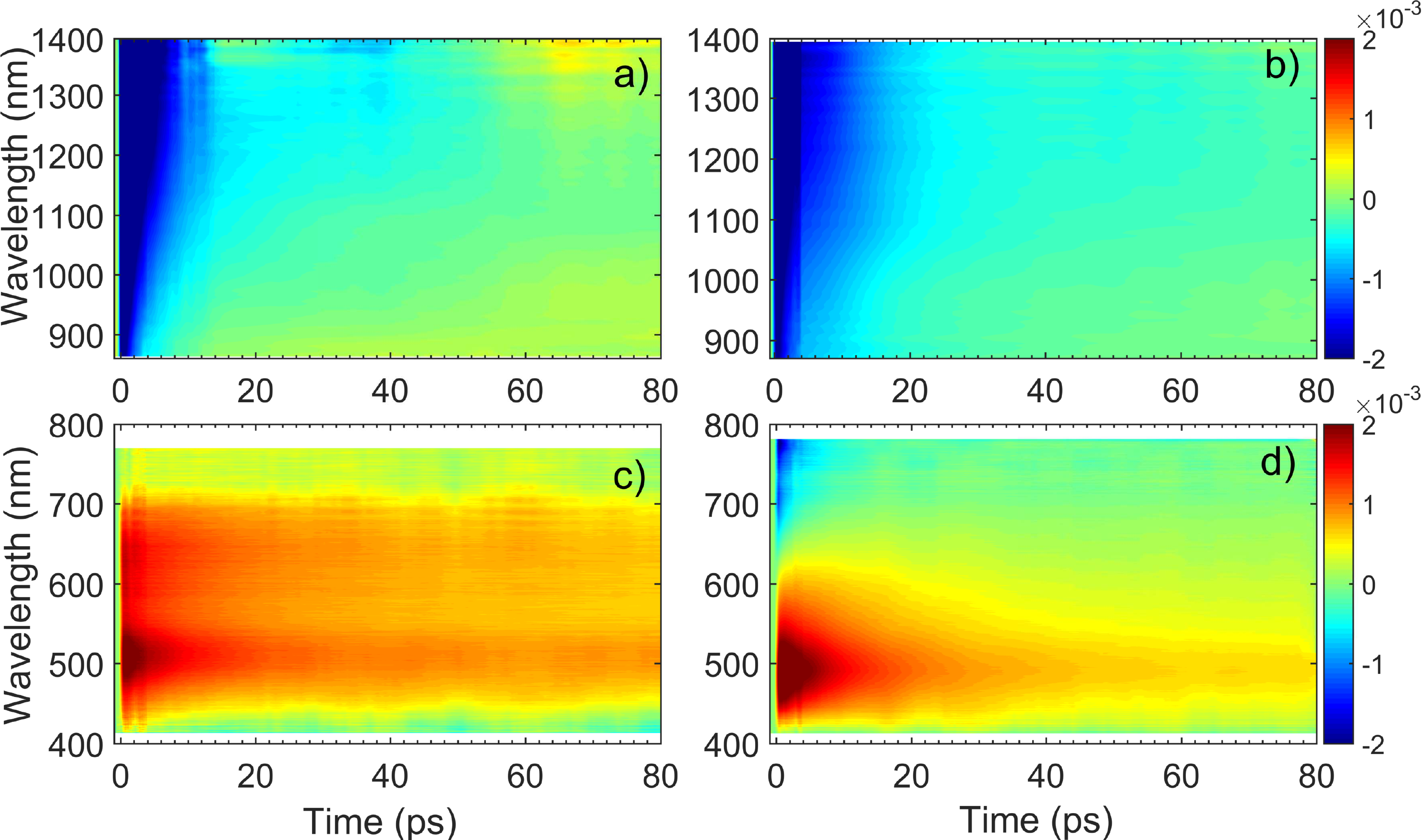}}
\caption{$\Delta$T/T maps as a function of probe wavelength and pump-probe delay of TiO${_2}$ in (a) NIR (c) visible. $\Delta$T/T maps of TiO${_2}$-Gr10:1 in (b) NIR and (d) visible}
\label{fig:Fig10}
\end{figure*}
\begin{figure*}
\centerline{\includegraphics[width=170mm]{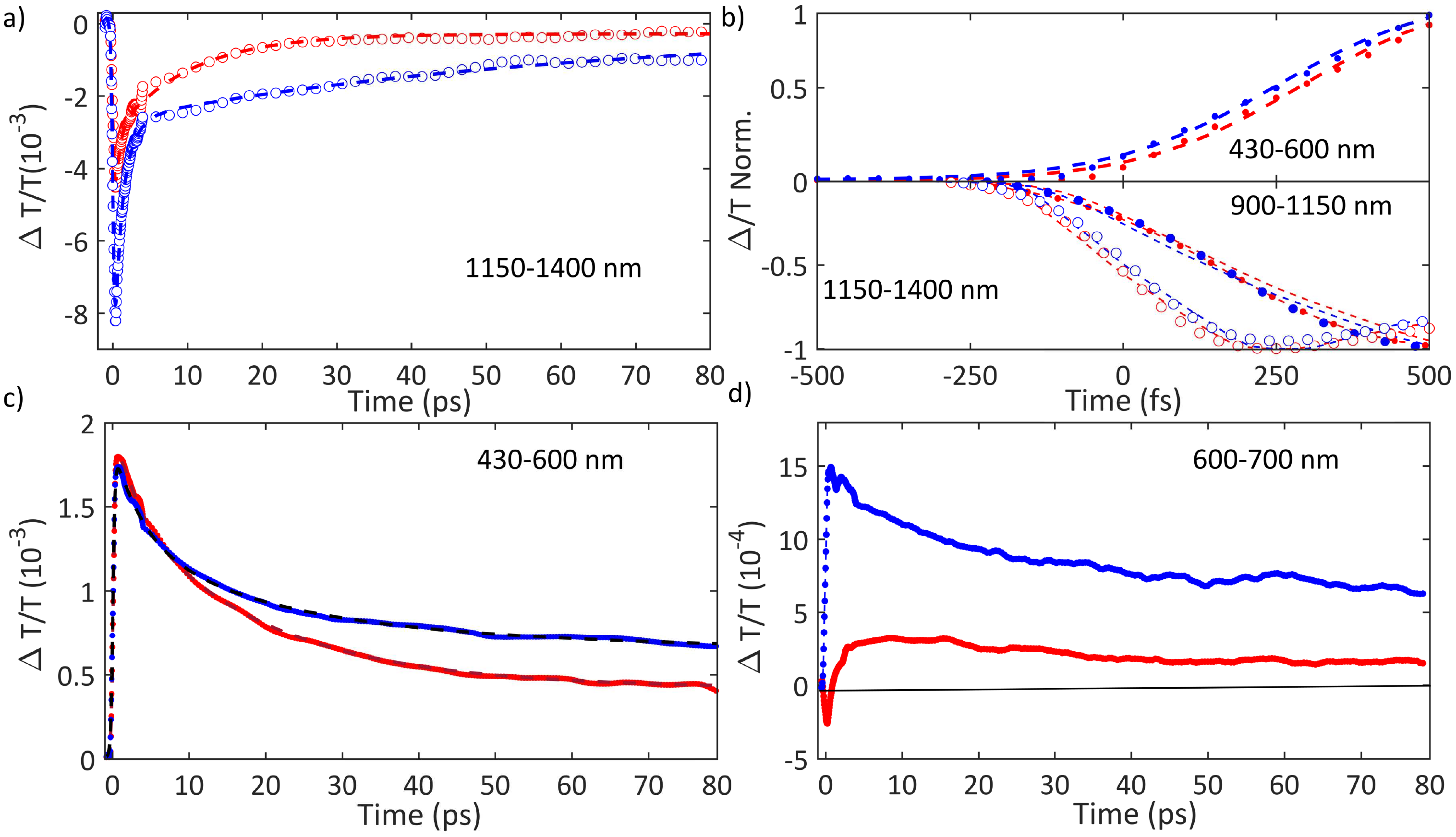}}
\caption{a,c,d) Relaxation dynamics of TiO${_2}$ (blue dots) and TiO${_2}$-Gr10:1 (red dots) together with the best fit (dashed lines). b) Normalized signal build up dynamics for the different transient transmission bands: SE in the range 430-600nm, PA in the range 1150nm-1400nm (open dots) and 900-1150nm (full dots).}
\label{fig:Fig11}
\end{figure*}
\begin{figure}
\centerline{\includegraphics[width=90mm]{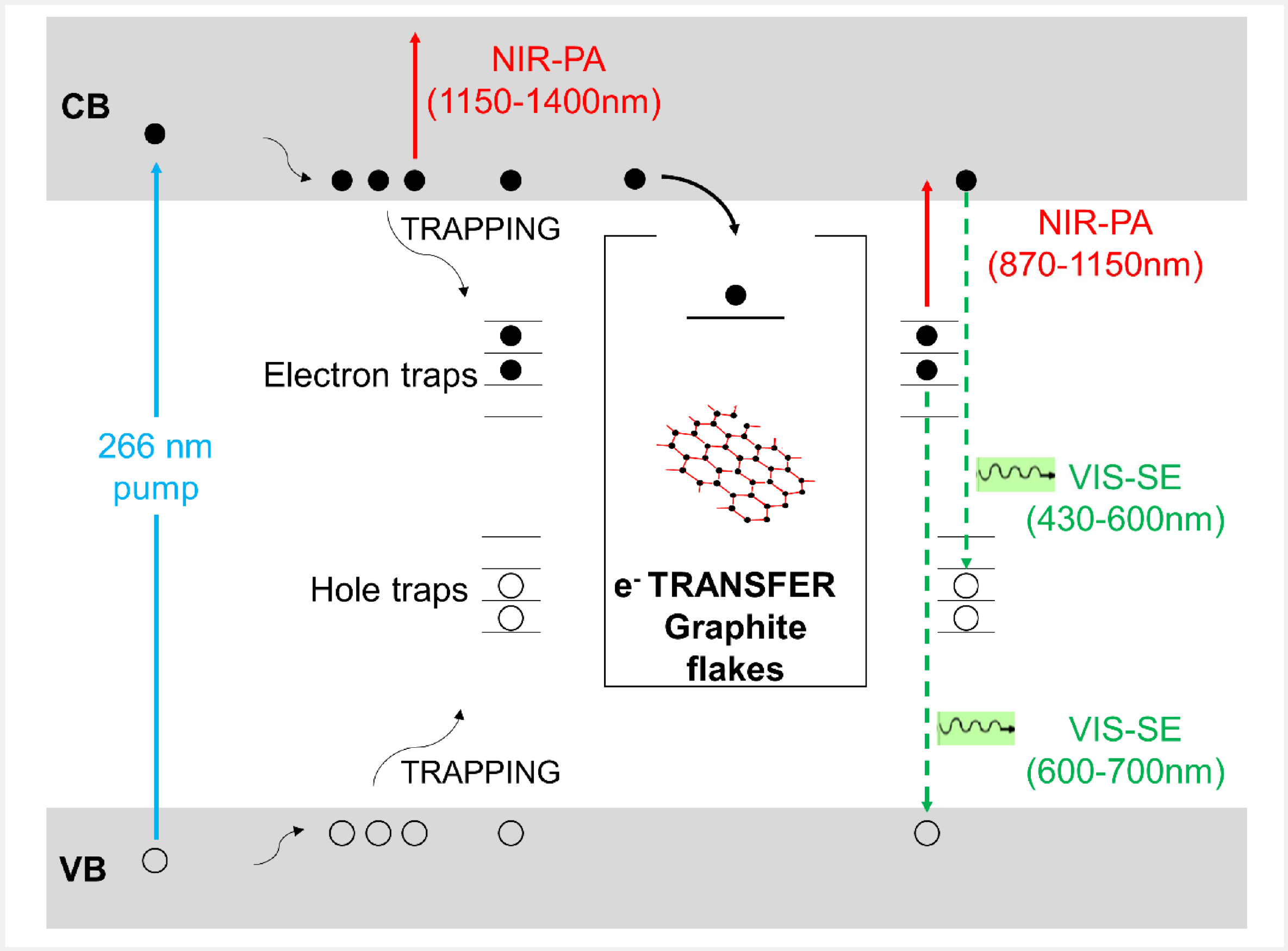}}
\caption{Schematic illustration of the optical transitions contributing to the TA signals of TiO$_{2}$ and TiO$_{2}$-Gr10:1. The UV-pump pulse (ciano arrow) photo-excites free-e (full black circles) and -h (empty circles) into CB and VB respectively. Photo-excited charge carriers thermalize to the bands edges and may also get trapped into inside-gap states. Photo-excited free-e are monitored by the probe pulse through the PA band in the range 1150-1400nm (red arrow) and the SE band in the range 430-600nm (green dashed arrow) related to the recombination with trapped holes. In TiO$_{2}$-Gr10:1, free-e can also transfer to graphitic flakes (inset box). The trapped electrons can radiatively recombine with free holes giving rise to the SE band in the range 600-700nm (green dashed arrow) or be photo-excited into CB as indicated by the PA bands in the range 870-1150nm.}
\label{fig:Fig12}
\end{figure}

Fig.\ref{fig:Fig7}a plots the UV-Vis absorbance, -log$_{10}$(T), with T the transmittance of TiO$_2$ and TiO$_2$-Gr10:1. These have two bands in the UV region at 270 and 306nm, characteristic of TiO$_2$-NPs\cite{Gu2008}, assigned to the first allowed vertical transitions that occur at the center of the Brillouin zone of TiO$_2$-NPs\cite{Serpone_2}. The PL spectra of TiO$_2$ and TiO$_2$-Gr10:1 in the liquid phase, following excitation at 266nm, are reported in Fig.\ref{fig:Fig7}b. While the shape of the spectra is similar, the PL intensity quenching in TiO$_2$-Gr10:1 points to an interaction between the excited TiO$_2$ and the exfoliated graphite, which prevents radiative recombination of the photogenerated e-h pairs.

The generation of reactive oxygen species\cite{Hirakawa_1} (ROS) was identified as the photodegradation mechanism of organic pollutants\cite{ZhangY2010} and RhB\cite{Gao2012}. The presence of exfoliated graphite in TiO${_2}$-Gr10:1 may result in a higher ROS generation, due to e-transfer from TiO$_2$ to graphite, allowing a more stable charge separation in TiO$_2$. The first step of the photocatalytic degradation reaction is the photo-excitation of e-h pairs in TiO$_2$-NPs by absorption of UV photons with energy exceeding the TiO$_2$ gap. The ROS generation depends on the competition between charge recombination, either radiative or non-radiative, and the separation of the photo-excited charges required to initiate the oxidative (reductive) pathways\cite{Hirakawa_1}. Accordingly, the enhancement of photocatalytic activity may be traced back to modifications of the relaxation channels of photoexcited e-h in TiO$_2$, induced by graphite flakes.

In order to identify these channels, we perform a comparative study of charge-carriers dynamics in pristine TiO${_2}$ and TiO${_2}$-Gr10:1 using broadband \textit{TA} spectroscopy with sub-200fs time-resolution. We use an amplified Ti:sapphire laser (Coherent, Libra) with 100fs, 500$\mu$J pulses at 800nm and 1kHz. The 266nm-pump pulse is generated by frequency tripling the laser output and it is modulated with a chopper at 500Hz. The broadband probe pulse is obtained by white light continuum generation in a plate of sapphire, for the visible, or yttrium aluminium garnet (YAG), for the near-infrared (NIR). The probe spectrum is detected by an optical multichannel analyzer with a wavelength resolution$\sim$1nm. The parallel linearly polarized pump and probe pulses are focused on the sample in a non-collinear geometry with spot sizes$\sim$180 and$\sim$80$\mu$m, in order to guarantee homogeneous excitation of the detected sample region. The pump power is 1.6mW, corresponding to an incident fluence$\sim$3mJ/cm$^2$ ($\sim$10$^{16}$ photons cm$^{-2}$). The measured signal is the delay-dependent differential transmission spectrum\cite{DeSilvestri2017}, defined as $\Delta T/T (\lambda,\tau$)=T$_{on}$($\lambda$, $\tau$)/T$_{off}$($\lambda$)-1, where T$_{on}$ and T$_{off}$ are the probe spectra transmitted through the excited and the unexcited sample, respectively, $\lambda$ is the probe wavelength and $\tau$ the pump-probe delay, controlled with a motorized translation stage. The temporal resolution is$\sim$180fs. We excite with UV pulses at 266nm, well above the band gap of TiO${_2}$\cite{Hoffmann1995}, and measure $\Delta T/T$ from 430 to 1400nm.

Fig.\ref{fig:Fig10} plots $\Delta T /T$ ($\lambda,\tau$) maps as a function of $\lambda$ and $\tau$. In the NIR, Figs.\ref{fig:Fig10}a,b, both TiO${_2}$ and TiO$_2$-Gr10:1 exhibit broad photo-induced absorption (\textit{PA}, $\Delta T /T<$0) from 870 to 1400nm. We assign it to intraband transitions of the photo-excited free e from the conduction band (CB) edge, as reported for anatase TiO${_2}$-NPs\cite{Szczepankiewicz2002,Yoshihara2004,YamakataA2001,Yamakata2001}. An additional source of \textit{PA} in the NIR comes from the transition of trapped e to the CB\cite{Knorr2008}. A large variety of trapping states is expected in TiO${_2}$, with energy distribution dependent on sample preparation\cite{Schneider2014}. According to Refs.\onlinecite{Yoshihara2004,Tamaki2006} the contribution of trapped e to the PA signal should dominate in the range 870-1150nm, while the free e absorption, which scales as $\lambda^n$ with n=1.7\cite{Yoshihara2004}, dominates at longer wavelengths. In the NIR, both SLG\cite{Breusing2011} and multilayer graphene\cite{Sun2008} show a positive $\Delta T /T$, corresponding to photo-bleaching \textit{(PB)} due to Pauli blocking\cite{Breusing2011,Sun2008} from the hot e distribution in the CB\cite{Brida2012,Tomadin2013}. Since the TA spectrum of TiO${_2}$-Gr10:1 in the NIR consists of a PA band, we conclude that its optical response is dominated by TiO${_2}$, due both to the higher intensity of the transient signal from TiO$_2$ and to the higher concentration of TiO$_2$ with respect to graphite flakes.

The TA maps of TiO${_2}$-Gr10:1 and TiO${_2}$ in the NIR differ for their time decay, as shown in Fig.\ref{fig:Fig11}a. The portion of the PA band in the range 1150-1400nm can be attributed to free e, as confirmed by the resolution limited formation of the signal in Fig.\ref{fig:Fig11}b, and by the monotonic increase of the signal with probe wavelength. For TiO${_2}$, this PA relaxes following a bi-exponential decay with time constants $\tau_{1\,TiO_2/PA}=500fs$, $\tau_{2\,TiO_2/PA}=45ps$. In the presence of exfoliated graphite, the relaxation dynamics is best fit by a three-exponential decay with time constants $\tau_{1\,G/PA}=500fs$, $\tau_{2\,G/PA}=4ps$, $\tau_{3\,G/PA}=20ps$. In both TiO${_2}$ and TiO${_2}$-Gr10:1, the first sub-ps decay component is associated to the trapping of free e\cite{Tamaki2006,Iwata2004,Tamaki2007}. The appearance of an additional decay channel, and the overall shortening of the PA bands lifetime observed in the composite with respect to the pristine TiO$_2$-NPs can be explained by ultrafast charge transfer from TiO${_2}$ to the graphite flakes, which act as e scavengers. The PA dynamics in the range 870-1150nm, mainly related to absorption from trapped e\cite{Yoshihara2004,Tamaki2006}, appears almost unperturbed by the presence of exfoliated graphite, suggesting that e transfer mostly involves free e. In both samples, this PA band shows a build-up with a 400-500fs time constant (Fig.\ref{fig:Fig11}b), related to e trapping. This rise time, consistent with the$\sim$200fs time constant measured in Pt-loaded TiO${_2}$ particles\cite{Furube2001}, matches the sub-ps decay component (indicated as $\tau_{1\,TiO_2/PA}$, $\tau_{1\,G/PA}$) of \textit{PA} in the range 1150-1400nm, observed in both TiO${_2}$ and TiO${_2}$-Gr10:1, which we attribute to free e trapping. Further evidence of e transfer from TiO${_2}$ to flakes can be found in the out-of-equilibrium optical response in the visible range, Figs.\ref{fig:Fig10}c,d. In the TiO${_2}$ sample we observe an increase in transmission ($\Delta T /T>$0) in the visible which, considering the vanishing ground state absorption in this spectral range, can be assigned to stimulated emission (SE), i.e. amplification of the probe beam due to optical gain\cite{DeSilvestri2017}. We identify two overlapping SE bands: the first, in the range 430-600nm, due to the recombination of free e with trapped h. The second, in the range 600-700nm, due to recombination of trapped e with free h. In TiO${_2}$-Gr10:1 the second, red shifted SE band is strongly quenched and a residual component appears few ps after excitation, Fig.\ref{fig:Fig11}d. The SE band in the range 430-600nm, related to trapped h recombination can be observed in both samples, but in TiO${_2}$-Gr10:1 it decreases faster to equilibrium, see Fig.\ref{fig:Fig11}c. This band has a single exponential build up with 400-500fs time constant, possibly due to h trapping, Fig.\ref{fig:Fig11}b. The SE relaxation dynamics can be fit by a bi-exponential decay on top of a long-lasting component related to the emission on the ns timescale\cite{Yamakata2001}. In TiO${_2}$, we get $\tau_{1\,TiO_2/SE}=5ps$, $\tau_{2\,TiO_2/SE}=45ps$, while in TiO${_2}$-Gr10:1 we have $\tau_{1\,G/SE}=4ps$, $\tau_{2\,G/SE}=20ps$ (Fig.\ref{fig:Fig11}c). While $\tau_{1\,TiO_2/SE}$ could depend on the lifetime of the trapped h, the other three relaxation components $\tau_{2\,TiO_2/SE}$, $\tau_{1\,G/SE}$ and $\tau_{2\,G/SE}$, match those observed for the PA decay in the NIR (equal to $\tau_{2\,TiO_2/PA}$, $\tau_{2\,G/PA}$, $\tau_{3\,G/PA}$) indicating that the SE band at 430-600nm and the PA band at 1150-1400nm decay with similar dynamics. These components can be associated to the population dynamics of free e, whose lifetime in TiO${_2}$-Gr10:1 is limited by the charge transfer to graphitic flakes, which occurs on a time-scale$\sim$4-20ps. Previous ultrafast spectroscopy  studies\cite{Tamaki2007,Furube2001} on Pt loaded TiO${_2}$-NPs suggested a similar e transfer time of several ps. In our case, e transfer to the graphitic flakes increases the trapped h lifetime, because it inhibits one of their recombination channels, enhancing the oxidative photocatalytic reactivity of the composite.

Fig.\ref{fig:Fig12} summarizes the photoexcitation and relaxation pathways of TiO$_{2}$ and TiO$_{2}$-Gr10:1 derived from our ultrafast TA experiments. In pristine TiO$_2$, the free e and h photo-excited into CB and VB by the UV-pump pulse (blue arrow), can either be excited by the probe pulse into higher energy states via intraband transitions responsible for the instantaneous PA band in the range 1100-1500nm, or they can relax into intragap trapped states. The trapped charge carriers can radiatively recombine with free charges giving rise to the SE bands in the range 430-600nm and 600-700nm. Trapped e can also be photo-excited into CB as indicated by the PA bands in the range 870-1150nm. All the bands related to the relaxation of trapped charge carriers share the same build-up dynamics due to the trapping. The interaction with graphitic flakes influences the optical properties of TiO$_2$-Gr, when compared to TiO$_2$, by opening an additional relaxation channel for the free e, which can efficiently transfer to the graphite flakes, thus slowing down e-h recombination, enhancing the photocatalytic activity.
\section{\label{Concl}Conclusions}
We reported TiO$_2$/Gr composites with enhanced photocatalytic activity with respect to pristine TiO$_2$-NPs. These are produced via liquid phase exfoliation of graphite in presence of TiO$_2$-NPs, without surfactants which could prevent the energy or charge transfer between TiO$_2$ and graphite flakes. The observed photo-degradation kinetics consists of a combination of zero-order and first-order processes. We assigned the increase in photocatalytic activity to electron transfer from TiO$_2$ to the graphite flakes, which occurs within the first ps of the relaxation dynamics. Due to the simplicity and cost effectiveness of the preparation procedure of our samples, we anticipate applications to smart photoactive surfaces for environmental remediation.
\section{Acknowledgments}
We acknowledge funding from the EU Graphene Flagship, EU Neurofibres, ERC Minegrace and Hetero2D, EPSRC Grants EP/509K01711X/1, EP/K017144/1, EP/N010345/1, EP/M507799/ 5101, and EP/L016087/1.

\end{document}